\documentclass{article}
\usepackage{arxiv}
\usepackage[utf8]{inputenc} 
\usepackage[T1]{fontenc}    
\usepackage{hyperref}       
\usepackage{url}            
\usepackage{booktabs}       
\usepackage{amsfonts}       
\usepackage{nicefrac}       
\usepackage{microtype}      
\usepackage{graphicx}
\usepackage{natbib}
\usepackage{doi}

\usepackage{amsmath}
\usepackage{caption}
\usepackage{subcaption}

\title{Microstructure Evolution of Solid Oxide Fuel Cell Anodes Characterized by Persistent Homology}

\author{ Piotr Pawłowski \\
	AGH University of Science and Technology\\
	Krakow, Poland \\
	\And
	\href{https://orcid.org/0000-0002-2959-3044}{\includegraphics[scale=0.06]{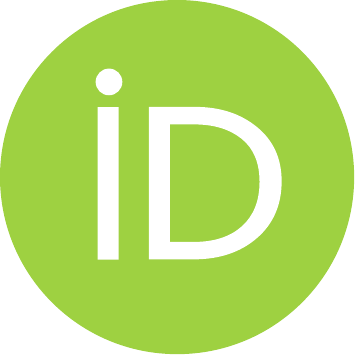}\hspace{1mm}Szymon Buchaniec} \\
	AGH University of Science and Technology\\
	Krakow, Poland \\
	\texttt{buchaniec@agh.edu.pl} \\
	 \And
	 \href{https://orcid.org/0000-0003-1898-571X}{\includegraphics[scale=0.06]{orcid.pdf}\hspace{1mm}Tomasz Prokop}\\
     AGH University of Science and Technology\\
	Krakow, Poland \\
	\texttt{buchaniec@agh.edu.pl} \\
	 \And
	 Hiroshi Iwai\\
    Kyoto University\\
	Kyoto, Japan \\
	 \texttt{iwai.hiroshi.4x@kyoto-u.ac.jp} \\
	 \And
	\href{https://orcid.org/0000-0003-4911-5880}{\includegraphics[scale=0.06]{orcid.pdf}\hspace{1mm}Grzegorz Brus \thanks{Corresponding author: Grzegorz Brus {brus@agh.edu.pl}}}\\
	AGH University of Science and Technology\\
	Krakow, Poland \\
	 \texttt{brus@agh.edu.pl}
}

\renewcommand{\shorttitle}{\textit{arXiv} Template}

\hypersetup{
pdftitle={A template for the arxiv style},
pdfsubject={q-bio.NC, q-bio.QM},
pdfauthor={David S.~Hippocampus, Elias D.~Striatum},
pdfkeywords={First keyword, Second keyword, More},
}

\begin{document}
\maketitle

\begin{abstract}
	Uncovering microstructure evolution mechanisms that accompany the long-term operation of solid oxide fuel cells is a fundamental challenge in designing a more durable energy system for the future. To date, the study of fuel cell stack degradation has focused mainly on electrochemical performance and, more rarely, on averaged microstructural parameters. Here we show an alternative approach in which an evolution of three-dimensional microstructural features is studied using electron tomography coupled with topological data analysis. The latter produces persistent images of microstructure before and after long-term operation of electrodes. Those images unveil a new insight into the degradation process of three involved phases: nickel, pores, and yttrium-stabilized zirconium. 
\end{abstract}


\section{Introduction}

A Solid Oxide Fuel Cell consists of solid porous electrodes separated by ceramic electrolyte. The electrodes are composed of a ceramic metal composite, in which every material has a specific mass or charge transport function, such as electron conduction, ion conduction, gas diffusion. The pathways for mass and charge flow are not straightforward, as they are dependent on the microscopic arrangement of conducting materials and open pores. Furthermore, the electrode-specific half reaction may only occur if reaction sites, such as triple phase boundaries or double phase boundaries are available. The overpotential generated in the microstructure accounts for a significant part of the voltage losses generated in an SOFC device. Currently microstructure-scale modeling relies on direct association of the electrochemical performance to physical attributes of a microstructure, such as phase volume fraction and phase tortuosity, grain size, availability of reaction sites. 
Microstructure-oriented analysis is generally based on a material reconstruction - a digital or analytical model of a micro-scale representative volume within the studied material. The digital material reconstruction may be based on either experimental measurements obtained using microscopic techniques, or an artificially generated model i.e. a collection of packed primitives, such as spheres. In the case of SOFCs the latter include stochastic microstructure reconstruction methods such as sphere-packing simulation \cite{schneider2006packing, nishida2011packing}, in which the electrode manufacturing process (powder mixing and sintering) is approximated. When a reconstruction is available, the properties of a microstructure may be approximated using experimental measurements complimented by analytical methods. 
stereological methods which extrapolate the three-dimensional distribution of phases in the microstructure from a limited number of cross-sections obtained by means of scanning-electron microscopy (SEM) \cite{yeong1998stereo}. Recently, advancements in tomographic techniques, such as X-Ray tomography, Focused Ion Beam Scanning Electron Microscopy, and their derivatives (i.e Plasma Focused Ion Beam SEM, Broad Ion Beam SEM) have allowed for preparation of three-dimensional digital reconstructions of SOFC microstructure \cite{wilson2006fibsem} corresponding directly to the distribution of material in a sample. The microstructure reconstruction may then be used to prepare a computational domain for solving mass and charge transport equations \cite{suzue2008sim, kanno2011simulation}. Alternately, the principles of the continuous electrode theory \cite{newman1975continuous} may be employed, and the electrode may be modeled as continuous medium, in which all the transport processes occur in parallel. The microstructure properties are then accounted for using macroscopic parameters, such as phase volume fraction, phase tortuosity, reaction site density \cite{iwai2010params}. The microstructure reconstruction may be used to estimate the aforementioned parameters: the volume expansion method \cite{iwai2010params}, the midpoint method may be used to estimate triple phase boundary density \cite{kanno2011simulation}, the marching cubes method  may be used to estimate interphase surface area \cite{lorensen1987marching}, diffusion simulation methods, such as the random walk method may be used to estimate sample tortuosity \cite{iwai2010params}.  
Alternately, empirical correlations may be implemented to estimate the transport parameters using grain size and volume fraction. However, the assumptions at the foundations of these methods may render them inaccurate in many applications \cite{tjaden2018tortuosity}.


The disadvantages of the quantitative approach related to the complicated process of obtaining data, together with the significantly limited possibilities of observing direct changes in the structure of the topological microstructure, present the necessity to propose an alternative method of describing the electrode material. Thanks to the use of tomography techniques to represent the SOFC anode material, it is possible to apply topological data analysis (TDA) methods, in particular persistent homology (PH). This solution allows for the observation of the characteristic properties of a given scale of the analyzed object and presenting them in the form of a persistence diagram (PD) or after appropriate data processing as a persistence image (PI). Persistent homology is a relatively young and still under development topological tool \cite{Edelsbrunner}, but it has already found application in many fields of science and engineering. 
TDA methods are particularly popular in medical research. The SARS-Cov-2 virus analysis by \cite{chung2021lattice} used persistent homology to represent the topological properties of the closed and open form of the coronavirus and to compare the differences between human and feline virus. The results of the research presented in the form of a persistence diagram allowed for the identification of characteristic properties for individual variants of SARS-Cov-2, which, according to the authors, may allow for the development of a faster and more accurate automated method for recognizing a given type of virus in the future. No less important analyzes based on TDA methods are also performed in other branches of medicine, i.e. neurobiology. In the publication of \cite{bendich2016persistent} researchers used persistent homology to describe the aging changes of brain tree arteries. On the basis of the obtained results, it was observed that even a significant filtration of the PH data allows for a satisfactory description of the aging factor impact. Research using persistent homology has found application in hepatocellular ballooning in liver biopsies analyzes (\cite{teramoto2020computer}), where persistence images were used, and in oncology (\cite{bukkuri2021applications}), allowing better observation of the cancer architecture. 
Another field of science that uses the development of TDA methods is geology. In the work of \cite{robins2016percolating} various two-phases systems were studied using persistent homology. These analyzes showed the correlation of the percolation threshold not only for 0th homology (connected components) and an improvement in determining the percolation radii was found if the data was previously filtered. The team of \cite{suzuki2020inferring} confronted the results obtained with topological data analysis with conventional methods, where in some cases the use of persistent homology provided additional information that was impossible to observe with the classical approach. 
TDA methods have also found application in flow analysis. In the research of the \cite{suzuki2021flow} authors used persistent homology, in particular 1st order homology (loops or open holes) to model physical phenomena occurring in porous material. Flow studies based on topological methods showed a high compliance with classical simulations and, importantly, there are no contraindications to apply this approach in the case of mass and energy transport processes. The possibilities of topological data analysis in the field of fluid mechanics were presented by \cite{kramar2016analysis} using persistence diagrams in both the vorticity during Kolmogorov flow and the Rayleigh-Bernard Convection temperature field studies. These analyzes confirmed the stability of persistent homology in the case of limiting the simulated values, which was observed by preserving the characteristic properties for the Rayleigh-Bernard Convection. 
TDA methods have found application in many other fields of science (\cite{app12010050, hirata2020structural, ICHINOMIYA20202926}), offering the possibility of insight into the structure of various systems. Based on the available scientific sources, it is clear that persistent homology seems to be the right tool for the analysis of the porous material of the solid oxide fuel cell anode. \par
This research aims to use computational topology in the new field, which is solid oxide fuel cell microstructure analysis. Apart from the new application of computational topology, we also use a unique set of data that includes a three-dimensional digital representation of anode microstructure before and after long-term operation. The research points to a new direction in fuel cell analysis and lays down the background to improve the utilization of deep learning for fuel cell study.   
\section{Background}
\begin{figure}[ht]
	\centering
	\includegraphics[width=0.6\textwidth]{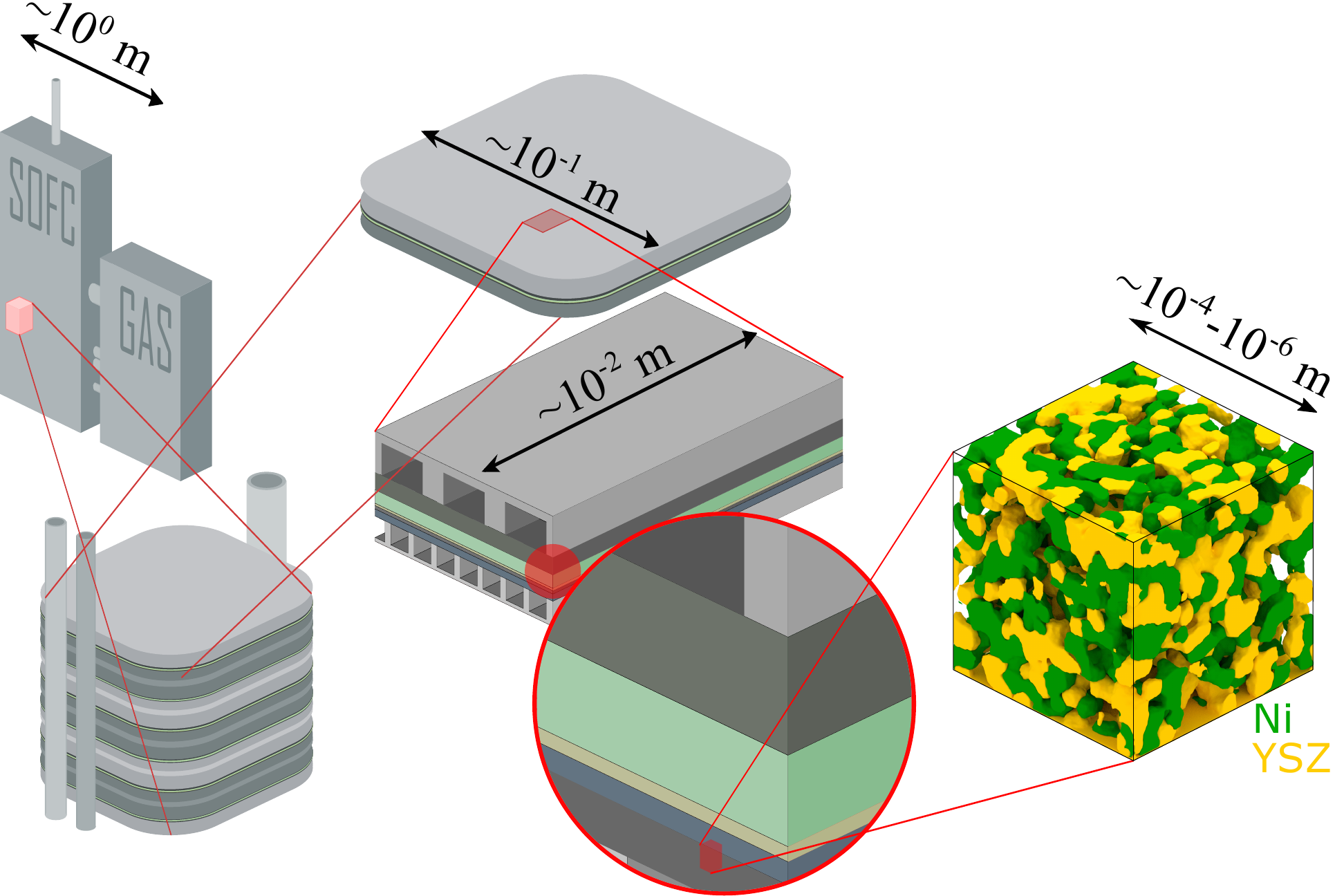}
	\caption{Modeling scales in an SOFC device}
	\label{modellevels}
\end{figure}

\begin{figure}[ht]
	\centering
	\includegraphics[width=0.6\textwidth]{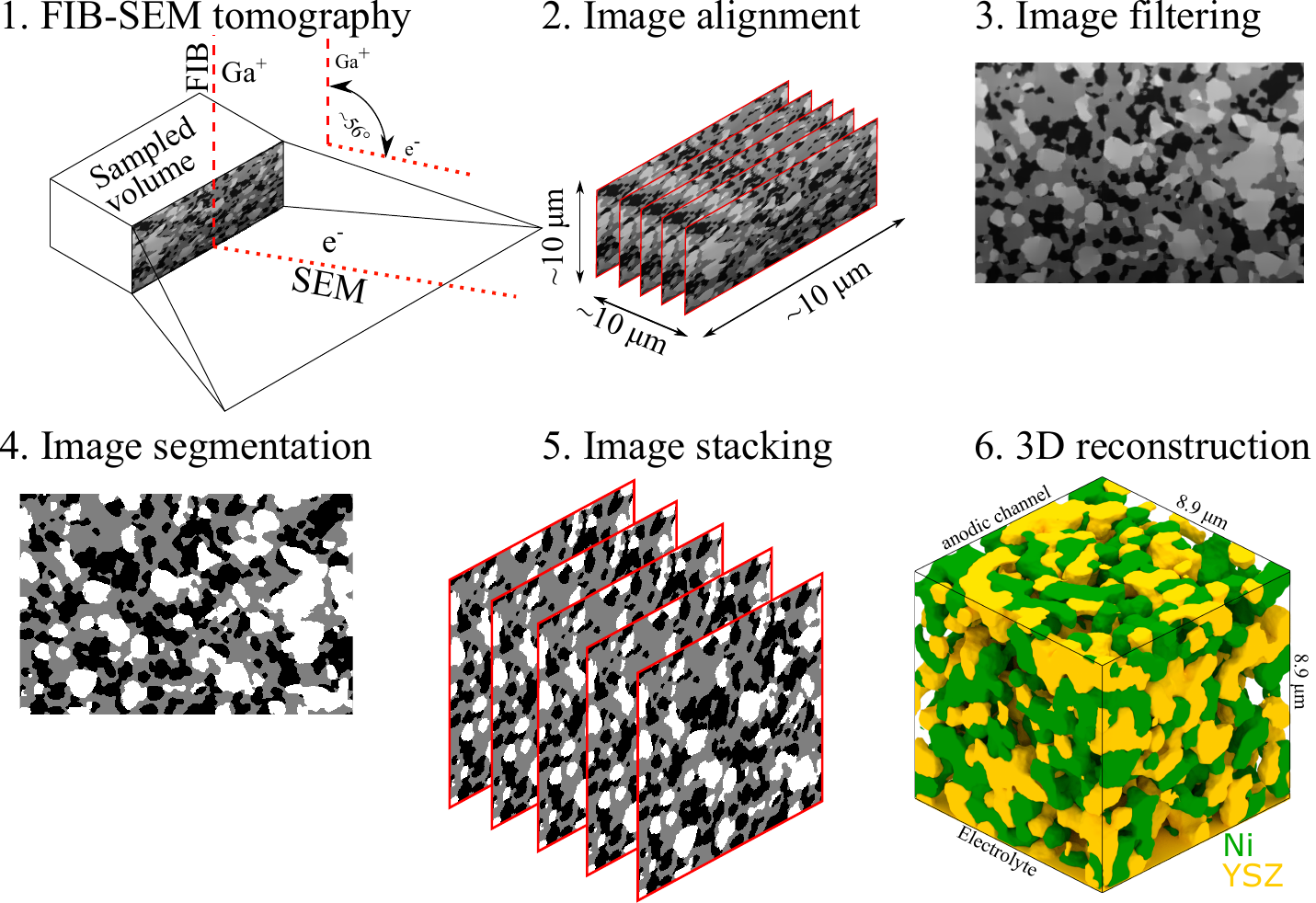}
	\caption{Focused Ion Beam Scanning Electron Microscopy in SOFC research}
	\label{fibsem}
\end{figure}

The efficiency of an electrochemical process is affected by the transport of reagents to and from the reaction sites. In the case of porous electrodes, the flow of of mass, charge and heat occurs within microscopic pores and grains of a ceramic metal or a ceramic-perovskite composite. Previously, we have employed microstructure-oriented modeling in a number of studies related to contemporary issues in Solid Oxide Fuel Cell development. The microstructure-scale in context of a typical SOFC device is illustrated in Figure \ref{modellevels}
Such an approach is important in the context of understanding the behavior of electrode material during long-term performance. The research subject has garnered a considerable interest of the scientific community, as many of the relevant physical processes, such as phase particle coarsening or impurity accumulation occur at the scale of the electrode microstructure.
In fact, the parameters obtained by quantifying Focused Ion Beam Scanning Electron Microscopy (FIB-SEM)-based digital reconstruction of the microstructure can be used to accurately predict SOFC stack electrochemical performance (see Fig. \ref{fibsem} for the framework of a typical FIB-SEM based microstructural analysis).
Phase volume fraction, phase tortuosity factor, and reaction site density (either the triple phase boundary, or the double phase boundary) are often a necessary and sufficient input for a mathematical model of voltage losses in a small SOFC device. This has been demonstrated in one of our studies, where a model using microstructural data was validated against the results of a 300 W SOFC stack experiment \cite{brus2017microstructure}. Multiscale model, such as the model developed and validated by Mozdzierz et al. \cite{mozdzierz2019multiscale} can further account for phenomena occurring apart from the microstructure, such as the irreversibilities of flow in an electrode channel. 
%
%
%

Previously, we have performed a number of solid oxide fuel cell long term performance studies, in which the emphasis was put on the microstructure evolution \cite{brus2015degradation, brus2015degradationtpb}. A 100 W 6-cell SOFC stack (an in-series or parallel connection of multiple cells), manufactured by SolidPower SPa of Italy, was operated over a period of 3700 h. The voltage generated by the system was being measured continuously, revealing not only a lack of deterioration of stack electrochemical parameters, but also a slight voltage increase of about 2.58\%. After the long-term operation experiment was completed, the stack was disassembled, and the cells were investigated using Focused Ion Beam Scanning Electron Microscopy (FIB-SEM). Similar measurements were performed on a reference cell (reference cell being a brand-new cell of the same composition, but one which did not participate in the long-term performance experiment). FIB-SEM study allowed us to obtain three-dimensional digital reconstructions the microstructures within the cells. The reconstructions were then analyzed using tomography data processing software (Avizo by Thermofischer Scientific/FEI) as well as in-house code to quantify the microstructural parameters such as Triple Phase Boundary (TPB) density and tortuosity of phases. 
The TPB density was determined to be a relevant parameter for the analysis, since it is the main reaction site within an SOFC anode, and its deterioration has often been perceived as one of the main causes of the SOFC degradation. On the other hand, tortuosity and volume fraction increase the resistivity to current conduction and gas diffusion. 
The studies have revealed  \cite{brus2015degradation, brus2015degradationtpb} that for all of the analyzed samples the TPB density decreased in relation to the values measured in the reference cell. But so did the tortuosity of the ion-conducting phase and the open pores. The degree to which the cells' microstructural parameters have changed were found to vary depending on the position within the stack. Additionally, the parameters of cells investigated after the long term performance experiment were found to exhibit higher anisotropy - the tortuosity measured in different directions varied more significantly for samples recovered after the stack experiment \cite{brus2020anisotropic}. Another long term stack experiment involving the SolidPower stack of the same type has shown a similar tendency.
In another study, the microstructure reconstruction was used to construct a three-dimensional computational domain for simulating microscale mass and charge transport in the electrodes \cite{prokop2019degradation}. Empirical relationships for material and substance parameters were used, along the common Bulter-Volmer model in order to solve the Poisson differential equation set for conservation of mass and charge. A numerical solution was obtained for microstructure samples taken from the reference cell, and the cells investigated after the degradation experiment. The numerical model predicted a decrease in voltage loss for the post-degradation samples based on their microstructures, thereby supporting the hypothesis that the increase of terminal voltage was the result of microstructure evolution.
The digital microstructure reconstructions were later used in a pilot study to determine the main cause of the microstructure evolution. By quantifying the interfacial free energy (a thermodynamic state function, which is expected to monotonously decrease for spontaneous processes) it was shown that the performance enhancement was likely the result of an expected nickel coarsening or a similar internal processes. \cite{prokop2021interfacial}.

\section{Mathematical model}
\begin{figure}[!th]
    \centering
    \includegraphics[width=0.6\textwidth]{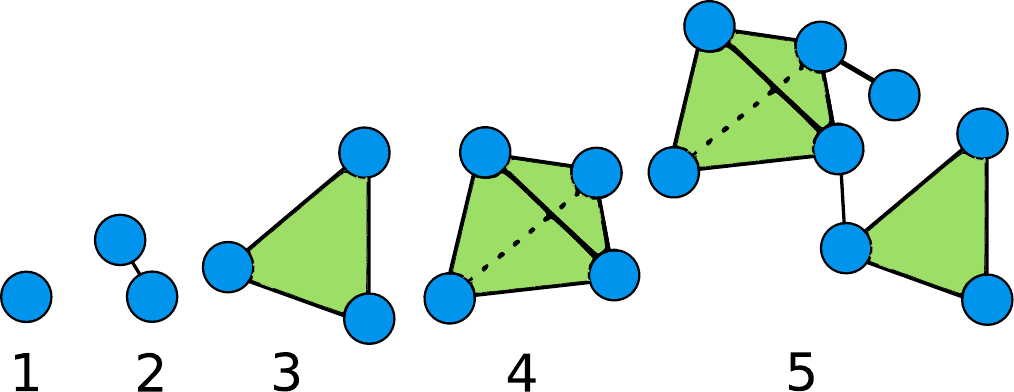}
    \caption{A representation of simplicial complexes. 1) vertex (0 - simplex). 2) segment (1 - simplex). 3) triangle (2 - simplex). 4) tetrahedron (3 - simplex). 5) Simplicial complex}
    \label{simplex}
\end{figure}
\begin{figure}[!ht]
    \centering
    \includegraphics{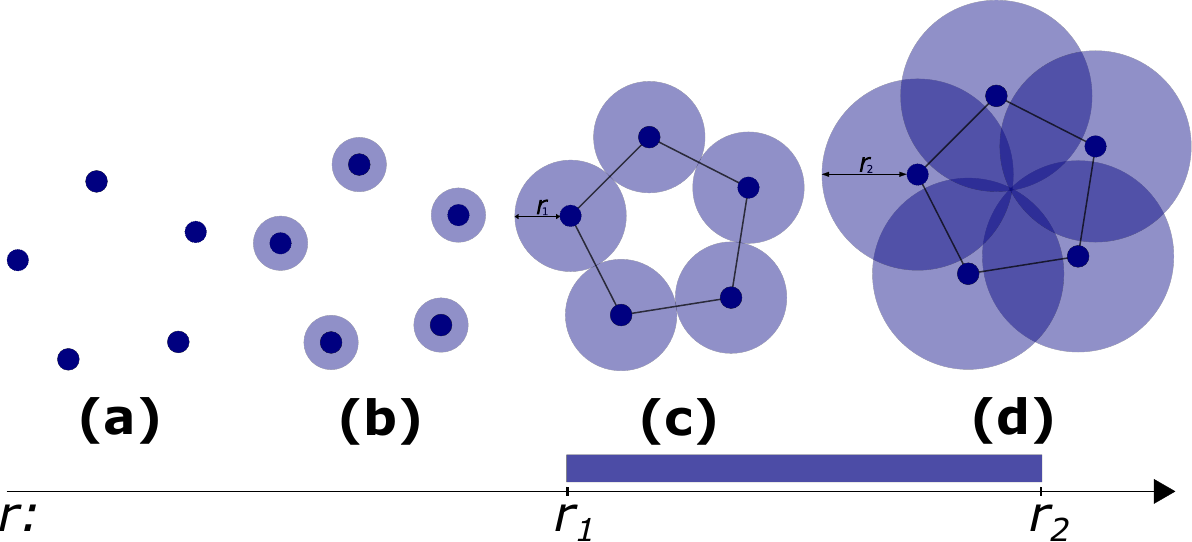}
    \caption{The formation of a simplicial complex based on the $r$-ball model. a) data set. b) radius increase. c) creation of a simplicial complex. d) giant simplex. The segment below the pictures shows the persistence of the topological feature with respect to the radius length}
    \label{barcode}
\end{figure}
In order to present the process of the creation of the persistence diagram, it is necessary to introduce a number of concepts related to the TDA methods. Homology is the topological invariant (a parameter that is immutable during the transformation of a given topological space) that represents a set of $n$-dimensional holes (\cite{obayashi2018persistence}). It characterizes the topological properties of a given scale of the analyzed object, allowing for comparison of two different topological spaces (\cite{pereira2015persistent}). The application of persistent homology requires the discretization of continuous space, which is performed by using the so-called complexes. There are many different types of methods used to represent a topological space. For the purpose of the research contained in this paper, the simplicial complexes are considered. This method consist in building cells based on simplices representing the simplest geometric objects with flat faces. A simplical complex $S$ should be understood as a set of simplices consisting of: points (0 - simplex), segments (1 - simplex), triangles (2 - simplex), tetrahedrons (3 - simplex) and other multidimensional simplices presented in Fig.\ref{simplex} (\cite{zhu2013persistent}) therefore, it can be concluded that for a given topological space $\mathbb{M}$, homology assigns a set of complexes $H_k(\mathbb{M})$ describing the topological properties of $\mathbb{M}$, where $k = 0, 1, \dots$ is the order of the homology. The dimension $H_k(\mathbb{M})$ is known as $k^{\mathrm{th}}$ Betti number and is commonly denoted as $\beta_k$ (\cite{edelsbrunner2013persistent}). The purpose of this parameter, which is a topological invariant, is to count the $k$-dimensional features of the studied space. Equivalent topological spaces have the same value of the Betti numbers. In this paper, the three-dimensional space will be examined, hence the $k$ = 0, 1 and 2 are considered, denoting connected components, loops (cycles) and voids. 
The main goal of persistent homology is therefore to calculate homology on many scales while observing topological properties of the system that persist across those scale. The term 'persistent' is associated with the appearance and vanishing of the observed properties of a given order of homology as a result of carrying out subsequent stages of topological space filtration. When considering a certain set of topological data (see Fig.\ref{barcode} (a)) in order to extract the characteristic properties of a system, complexes are constructed. For the purpose of filtering presented data set, the $r$-ball model was applied such that:
\begin{equation}
    T_r = \bigcup\limits_{i = 1}^{n} B_r(x_i),
\end{equation}%
where $T_r$ is the set of all points distant from any element of the $P$ by $r$, $B_r(x_i) = \{y \in \mathbb{R}^N: ||y - x_i|| \leq r\}$ is a ball centered at $x_i$ with radius $r$ and $N$ denotes the dimensionality of space \cite{obayashi2018persistence}. As the radius increases segments (1 - simplex) are created which delimit a certain two-dimensional space (Fig.\ref{barcode} (c)). Thanks to this process $\mathrm{PH}_1$ homology (cycle) is distinguished. Further radius enlargement results in a creation of a face (2 - simplex), which causes the loop to disappear completely (Fig.\ref{barcode} (d)). In order to determine the persistence of the topological properties, the whole process of space filtration is monitored where the lifespan of individual homology is described by a segment from birth to death (understood as appearance and disappearance of the feature). 

Constructing a persistence diagram from a barcode is relatively easy - instead of drawing a segment of the length from appearance of a topological property to its disappearance, a point with coordinates of birth and death is plotted on the graph. Therefore, persistence diagram $\mathrm{PD}_k$ should be understood as representation of the set of all $k^{\mathrm{th}}$ homology elements. Usually in the PD diagram x-axis corresponds to the birth of a given topological property, while the y-axis represents its death. A characteristic element of PD is the diagonal $f(x) = x$ dividing the plot area into to regions, where, due to the nature of data representation, the part located under the line is always empty (it is impossible for a topological property to die before its appearance). Points lying on the diagonal and in its direct proximity are categorized as noise (due to their negligibly short life), while those positioned at a considerable distance from function $f(x) = x$ represents proper topological properties. 

\begin{figure}[!ht]
\centering
    \begin{subfigure}{0.45\columnwidth}
        \centering
        \includegraphics{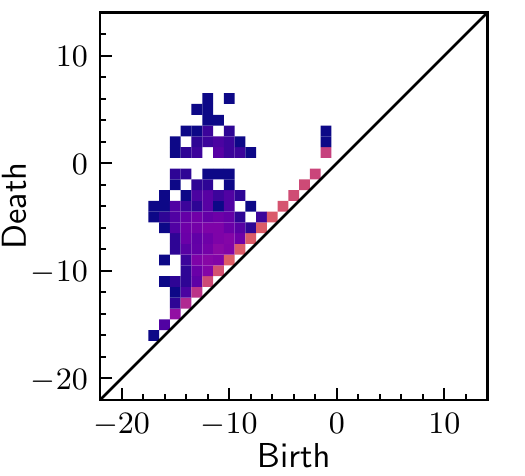}
        \caption{Original persistence diagram}
    \end{subfigure}
    \begin{subfigure}{0.45\columnwidth}
        \centering
        \includegraphics{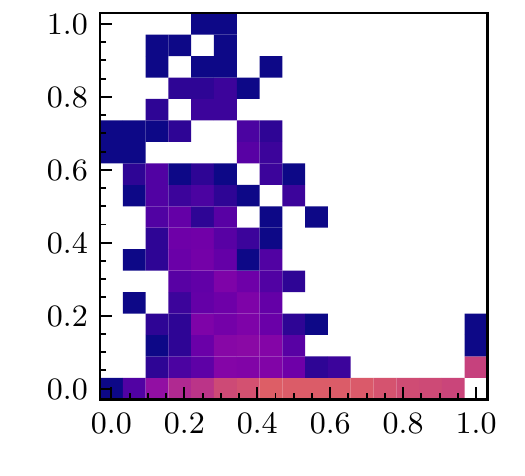}
        \caption{Linear transformed persistence diagram}
    \end{subfigure}

    \begin{subfigure}{0.45\columnwidth}
        \centering
        \includegraphics{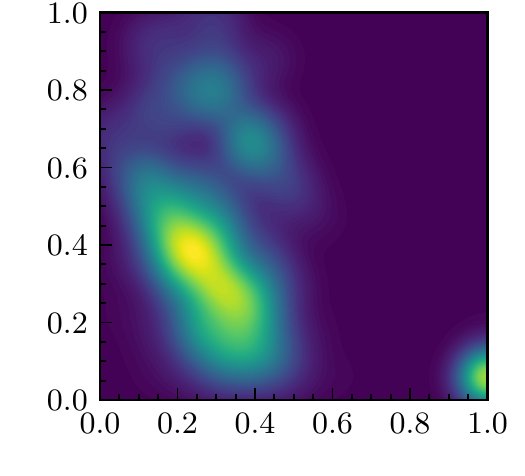}
        \caption{Persistence surface}
    \end{subfigure}
    \begin{subfigure}{0.45\columnwidth}
        \centering
        \includegraphics{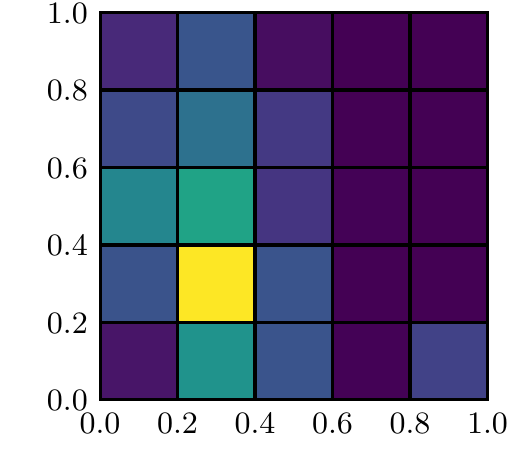}
        \caption{Persistence image}
    \end{subfigure}
\caption{Transformation of persistence diagram into a persistence image}
 \label{pd_to_pi}
\end{figure}
The representation of topological data in form of PD, although relatively convenient, does not always work well in further computer analyzes (\cite{adams2017persistence}). Therefore, there was a need to simplify the presentation of data used in subsequent stages of research based on methods , i.e. machine learning or decision tree algorithms. Such a solution appears in the so-called persistence image (PI), which is the conversion of the input PD to a finite-dimensional vector. This method not only retains the connection to the original persistence diagram and the topological properties it represents, but is also resistant to noise, efficient to compute, and has the ability to adjust the representation for a given system. The first step in converting the topological data representation of the system into PI is to perform linear transformation $N: \mathbb{R}^2 \rightarrow  \mathbb{R}^2$ of the original PD, such that: $N(x,y) = (x,y - x)$ \cite{adams2017persistence}. Then so-called persistence surface (PS) is created by applying a weighted sum of functions defined with a distribution $g_u$ centered at each point in the PD. In the research presented in this paper normalized symmetric Gaussian was used. For $g_u: \mathbb{R}^2 \rightarrow  \mathbb{R}$ with mean $m = (m_x, m_y)$ and variance $\sigma^2$, the $g_u$ is defined as \cite{adams2017persistence}:
\begin{equation}
    g_u(x,y) = \frac{1}{2\pi\sigma^2}e^{-\frac{(x-m_x)^2+(y-m_y)^2}{2\sigma^2}},
\label{gu}
\end{equation}%
and persistence surface $S_N$ as:
\begin{equation}
    S_N(x,y) = \sum\limits_{d \in N(x,y)} w(x,y)g_u(x,y).
\end{equation}%
The introduction of the weight function $w$ is due to the necessity to ensure the stability of the persistence surface. The last transformation leading to PI is done by integrating $S_N$ over sub-areas $p_i = {(x,y) : p_j\cap p_k=\emptyset, \bigcup_k p_k = \mathbb{R}^2}$ of a discretized domain, which can be represented as \cite{adams2017persistence}:
\begin{equation}
    I_{S_N}(p_i) = \iint_{p_i} S_N \mathrm{d}y\mathrm{d}x. 
\end{equation}%
The representation of topological data in form of persistence image is influenced by a number of parameters selected depending on the system under consideration. The important issue is the selection of the appropriate grid (often recognized in the literature as resolution) for the persistence image. Research shows (\cite{adams2017persistence}) that the data presented by means of PI relatively well preserve the representation of characteristic topological properties even for significantly reduced resolution. Another crucial aspect worth consideration is defining the appropriate weight function for a given problem. It was documented that the points with the longest persistence are not always the most essential in the analysis of a studied system (see \cite{bendich2016persistent}). This parameter should be selected based on the type of problem under consideration. Applying a different distribution is also possible and determining the correct value of the variance should be examined. Transformation from persistent diagram into persistent image are presented in Fig. \ref{pd_to_pi}. As persistnce landscape and image require additional parameters, that have to be selected for given application, in this work only persistence diagrams will be presented and compared.

\section{Numerical model}

\begin{figure}[!h]
	\centering
	\includegraphics[width=0.7\textwidth]{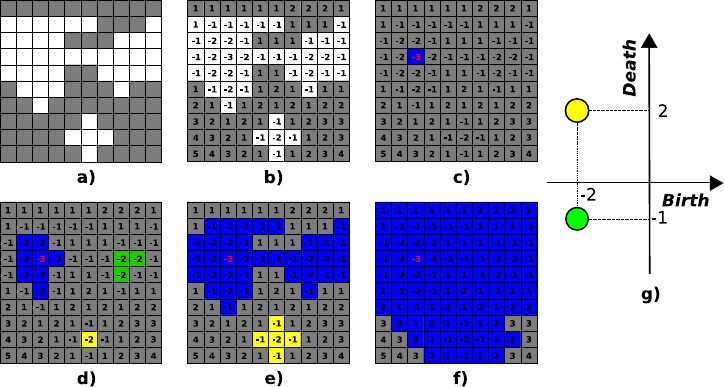}
	\caption{Connecting components. a.) Input data. b) Manhattan distance. c) The deepest filtration $r = 0$, the birth of the first component (blue). d) Second stage of filtration $r = 1$, the 
birth of the two components (green and yellow). e) $r = 2$ - death of green component. f) $r = 4$ - death of the yellow component. g) $0$-th order persistence diagram $\mathrm{PD}_0$}
	\label{numerical}
\end{figure}

In order to perform a topological analysis of the SOFC anode microstructures, it was decided to use the HomCloud software \cite{obayashi2022persistent}. This choice was dictated mainly by the possibility of using as input data the series of gray-scale images representing the microstructure, as well as documented application in broadly understood engineering field (\cite{ICHINOMIYA20202926, hirata2020structural, suzuki2021flow, suzuki2020inferring}). The software allows to carry out analyzes for connected components, cycles and voids. The results can be obtained both in the form of a persistence diagram and a text file containing the coordinates of the birth-death points. The algorithm that the HomCloud software is based on, for analyzing binary images applies a function that uses Manhattan distance to assign appropriate values to pixels (or voxels) depending on the position relative to the phase boundary (\cite{obayashi2018persistence}). In the example of $\mathrm{PD}_0$ formation presented in Fig. \ref{numerical} (based on \cite{obayashi2018persistence}), the gray phase has positive values and the white phase - negative values. Then, data filtration is performed, which is manifested by the reduction of the analyzed phase until the point of birth of the oldest component is determined. The next step is to carry out the reverse process, i.e. inflation of the studied phase until all components merge into one structure. The moments of connecting of individual components are associated with the death of the younger of them (in the case of topological properties in the same age, the choice of the vanishing component is unrestricted). The adopted numerical model results in reduced PD0, due to the fact that the oldest component never disappears. Note that due to lack of distance equal to 0, there will be no points on persistence diagram with birth or death equal with this value.
%
\section{Results}

\begin{figure}[!t]
    \centering
    \includegraphics[width=0.6\textwidth]{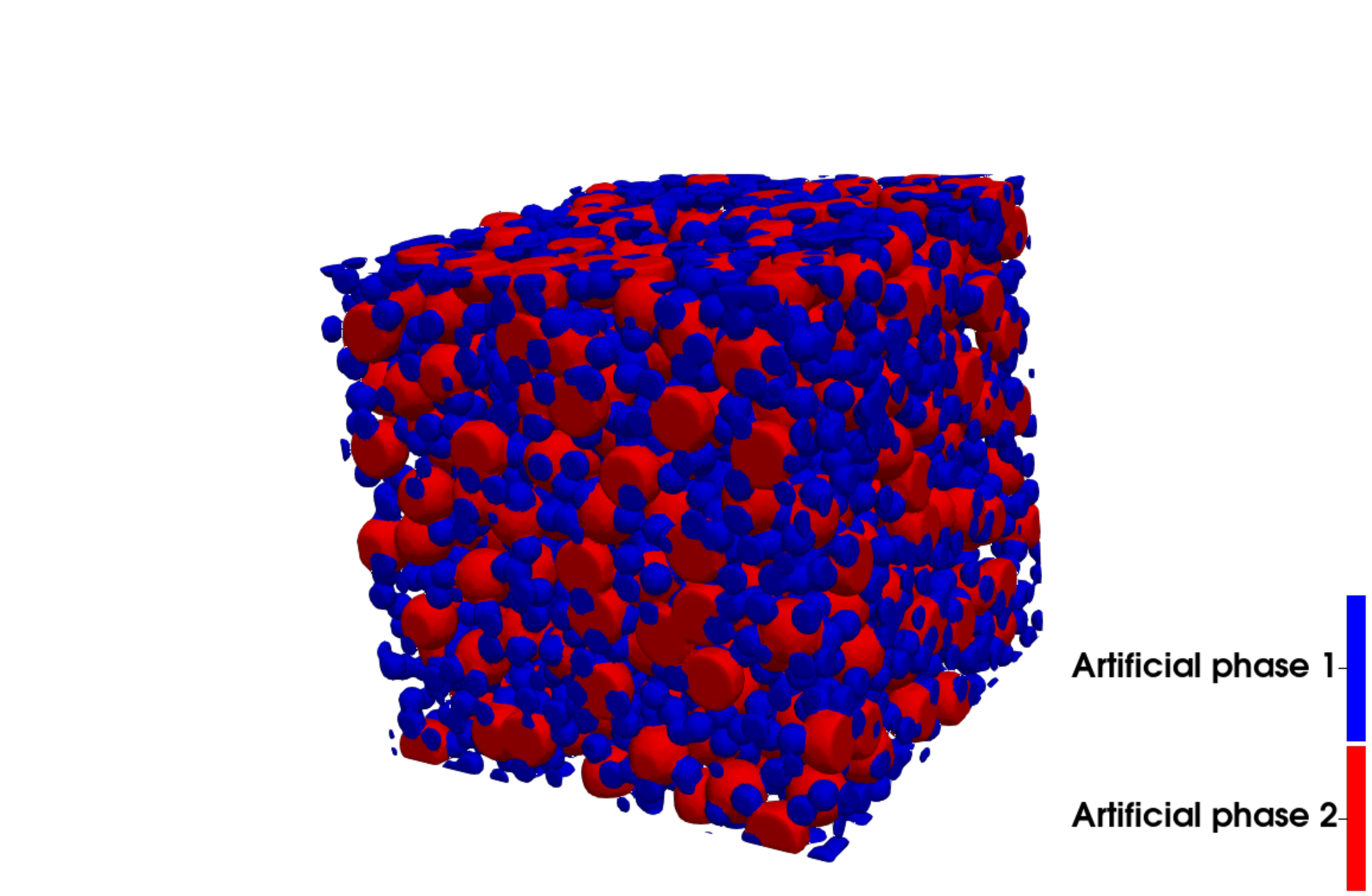}
    \caption{An example of an artificially generated synthetic microstructure composed of different spheres and used for the investigation shown in Fig. \ref{fig:test_series}. Blue and red spheres indicate artificially generated phase 1 and  phase 2 and void represents pores}
    \label{Artificial_microstructure}
\end{figure}

\newcommand{\widthThreeCol}{0.32}
\begin{figure}[!t]
\centering
    \begin{subfigure}{\widthThreeCol\textwidth}
        \centering
        \includegraphics{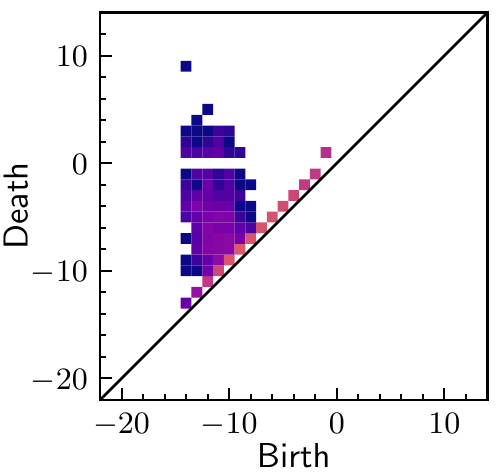}
        \caption{Realization No. 1, sphere diameter 1.0 ${\mathrm{\mu m}}$with one-point distribution }
    \end{subfigure}
    \begin{subfigure}{\widthThreeCol\textwidth}
        \centering
        \includegraphics{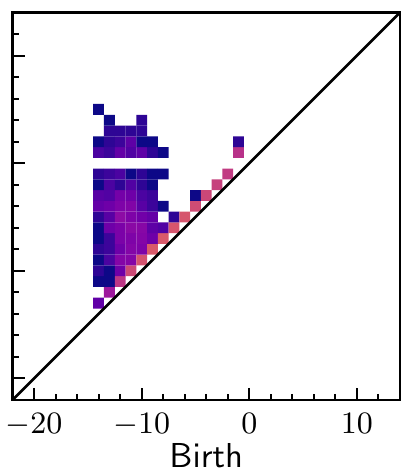}
        \caption{Realization No. 2, sphere diameter 1.0 ${\mathrm{\mu m}}$ with one-point distribution}
    \end{subfigure}
    \begin{subfigure}{\widthThreeCol\textwidth}
        \centering
        \includegraphics{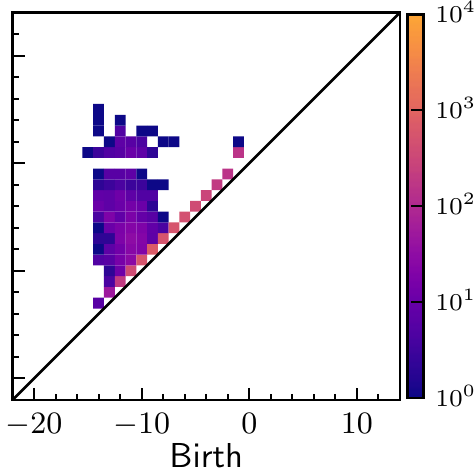}
        \caption{Realization No. 3, sphere diameter 1.0 ${\mathrm{\mu m}}$ with one-point distribution}
    \end{subfigure}
    \begin{subfigure}{\widthThreeCol\textwidth}
        \centering
        \includegraphics{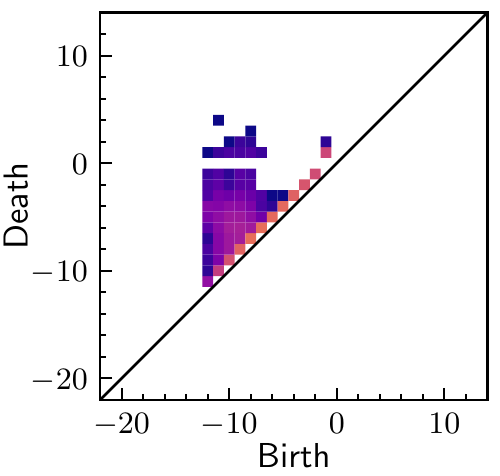}
        \caption{Sphere diameter 0.8 ${\mathrm{\mu m}}$}
    \end{subfigure}
    \begin{subfigure}{\widthThreeCol\textwidth}
        \centering
        \includegraphics{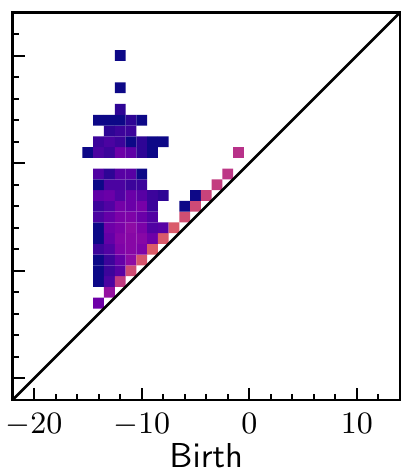}
        \caption{Sphere diameter 1.0 ${\mathrm{\mu m}}$}
    \end{subfigure}
    \begin{subfigure}{\widthThreeCol\textwidth}
        \centering
        \includegraphics{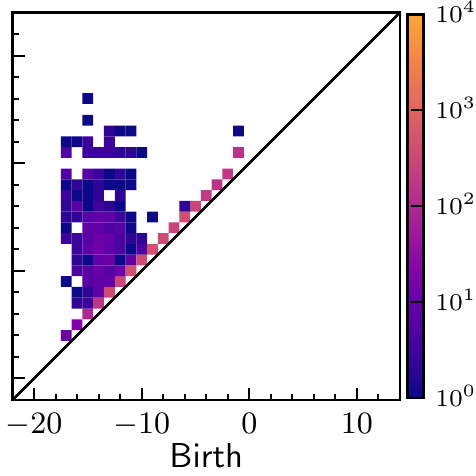}
        \caption{Sphere diameter 1.2 ${\mathrm{\mu m}}$}
    \end{subfigure}
    \begin{subfigure}{\widthThreeCol\textwidth}
        \centering
        \includegraphics{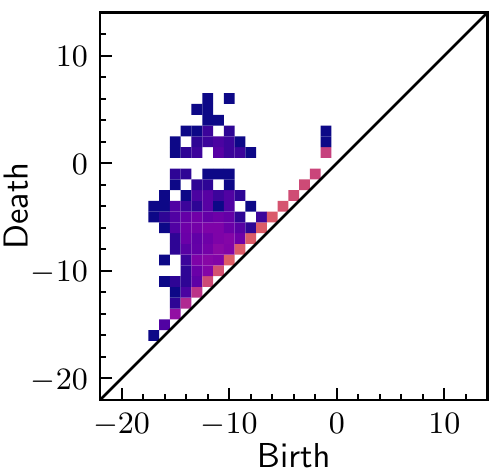}
        \caption{Standard deviation of sphere diameter 10\%}
    \end{subfigure}
    \begin{subfigure}{\widthThreeCol\textwidth}
        \centering
        \includegraphics{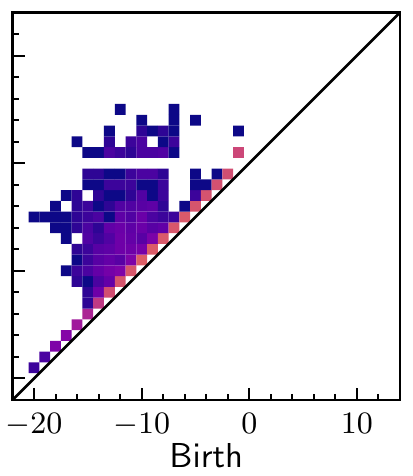}
        \caption{Standard deviation of sphere diameter 25\%}
    \end{subfigure}
    \begin{subfigure}{\widthThreeCol\textwidth}
        \centering
        \includegraphics{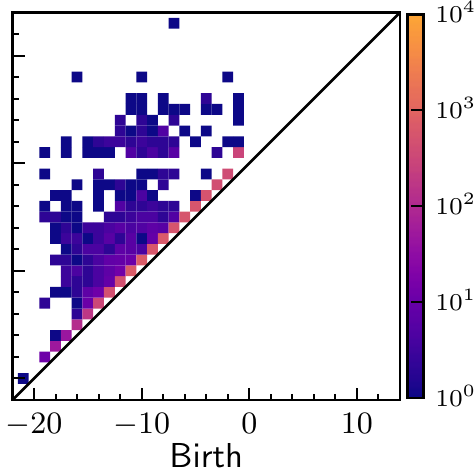}
        \caption{Standard deviation of sphere diameter 40\%}
    \end{subfigure}
\caption{$\mathrm{PD}_0$ diagrams from test series of consisting of varying synthetic microstructures. (a)-(c) different realizations of artificial microstructures, (d)-(f) realizations with varying sphere diameter (g)-(i) realizations with normal distribution of sphere diameters with varying standard deviation}
\label{fig:test_series}
\end{figure}
When applying the TDA methods to represent microstructural changes in anode material, a valid question arises about the accuracy of the technique and its sensitivity to changes during the aging process. While geological and, above all, material science studies using persistence diagrams are particularly valuable from the perspective of the SOFC fuel electrode, they do not fully answer the possible problems that should be considered when analyzing this material. The possible lack of projection of the loss of continuity of a given phase or the appearance of a new tunnel in the case of SOFC cells may be associated with a misrepresentation, resulting in erroneous predictions as to the operating parameters of the entire device. For these reasons, it was decided to conduct synthetic microstructure analyzes consisting of two phases in form of spheres and a third phase filling remaining space, as shown in Fig \ref{Artificial_microstructure}. Three series of tests were performed, analyzing the scale and significance of changes in the realization of the same random fields for system with the same parameters (volume fraction, particle size, sample size), with increasing the diameter of the spheres in subsequent samples and with the varying particle sizes inside the sample drawn from normal distribution. Selected $\mathrm{PD}_0$ and $\mathrm{PI}_0$  for the test series are presented in Fig. \ref{fig:test_series}. Research shows that the representation of synthetic microstructures is relatively stable to the influence of various factors. The differences for the first series are, as expected, minimal - there are minor disproportions between the diagrams (especially $\mathrm{PD}_0$), but the character typical for the order of the PD is retained in each case. It should be emphasized that the discrepancies are clearly understandable - they stem from the differences in the distribution of the spheres within the sample. The results obtained for second series, although more diversified, allow to distinguish even in a visual approach the direction of changes taking place in the microstructure while maintaining the general characteristics of the representation. It is evident that with the increasing radius of the spheres more persistent features appear. The greatest contrast between the representation is observable for the series that introduced varying particle sizes within the sample. However, even in this case, it is possible to distinguish the direction of increasing "dispersion" of birth-death points in relation to the reference case. The data representation in the form of PI also shows similarities and characteristic trends of topological changes even for third series. Nonetheless, it should be emphasized that selected linear weight function marginalizes distant single points of high persistence in favor of a large number of features in close proximity of diagonal, which results in less visible differences between images. The conducted tests for synthetic microstructures should not be extrapolated without reflection for real SOFC anode materials, but nevertheless they demonstrate the PD representation abilities and indicate possible directions of analyzes of the obtained data.

\begin{figure}[!h]
    \centering
    \includegraphics[width=0.25\textwidth]{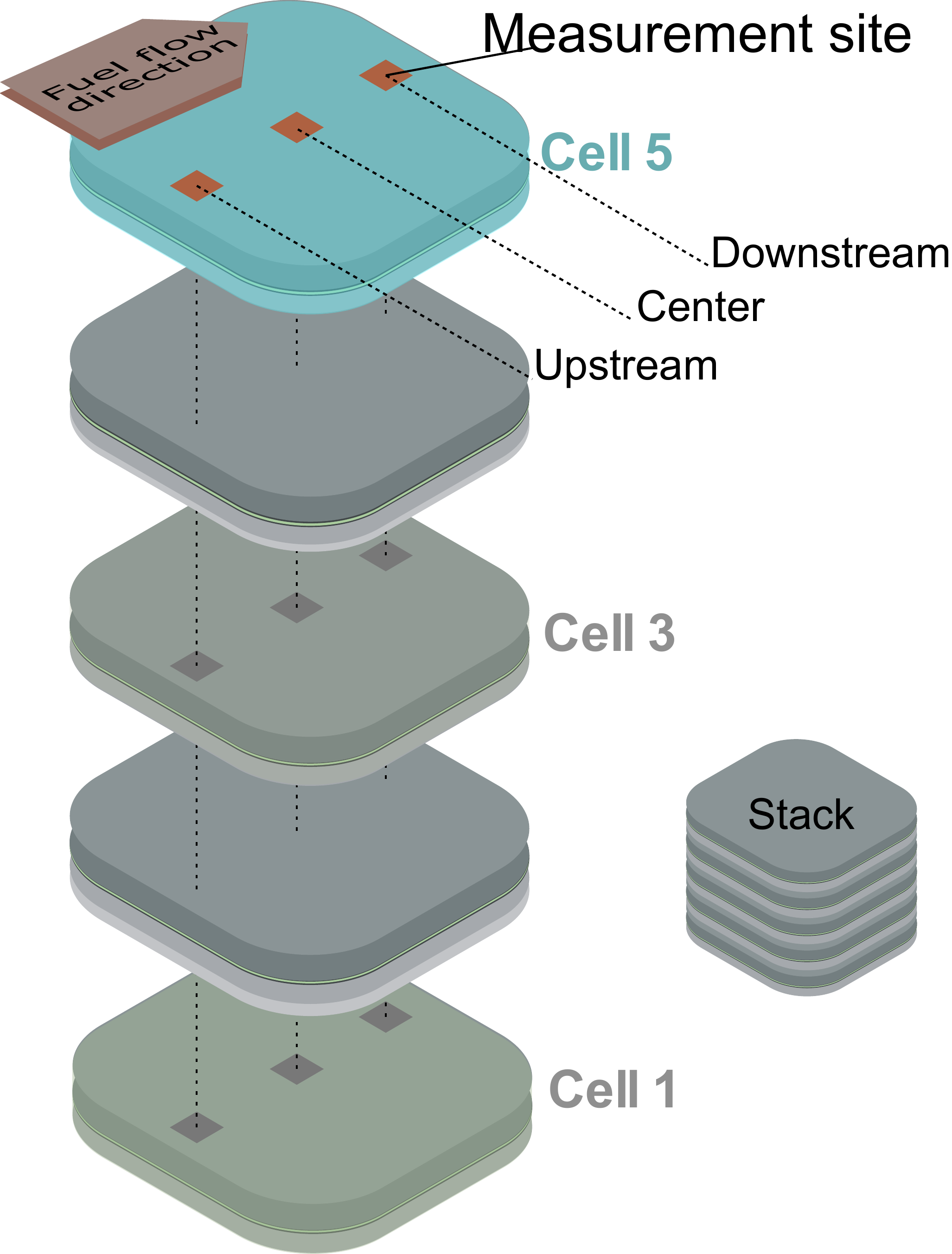}
    \caption{The sampled locations in the post-degradation stack}
    \label{fig:samplesites}
\end{figure}

The main goal of the research was to perform a topological analysis of SOFC anode microstructures belonging to devices in the stack subjected to the degradation process. The research used a stack manufactured by SOLIDPower S.p.A, analyzed in terms of quantitative approach in our previous works (\cite{brus2015degradation,brus2015degradationtpb,brus2020anisotropic,prokop2021interfacial}). The system consisted of six anode-supported cells with dimensions of 6 cm $\times$ 8 cm located in the modular stack test bench (MSTB). The stack was subjected to the degradation process by placing it in an furnace for 3800 h with properly selected thermal parameters. The anode material analyzed in this paper was taken from three locations in the cell recovered from the top of the stack: near the fuel inlet, near the fuel outlet, and in the center. 
Fig. \ref{fig:samplesites} shows a schematic of a stack with marked places from which the samples were obtained - the notation will be used in the following section to distinguish individual samples. In order to assess the scale of the changes taking place in the stack as a result of the aging process, a reference microstructure was taken from the brand new device.

\begin{figure}[!t]
\centering
    \begin{subfigure}{\widthThreeCol\textwidth}
        \centering
        \includegraphics{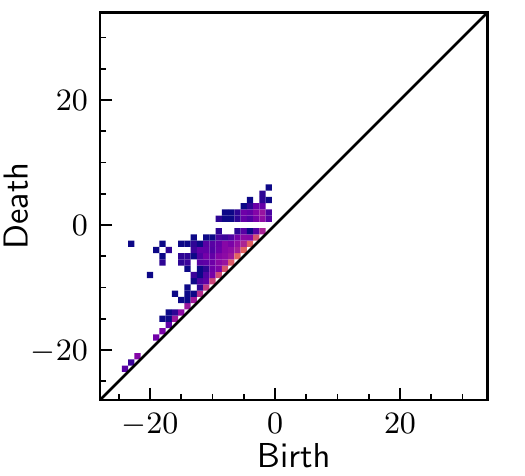}
        \caption{Nickel phase $\mathrm{PD}_0$}
    \end{subfigure}
    \begin{subfigure}{\widthThreeCol\textwidth}
        \centering
        \includegraphics{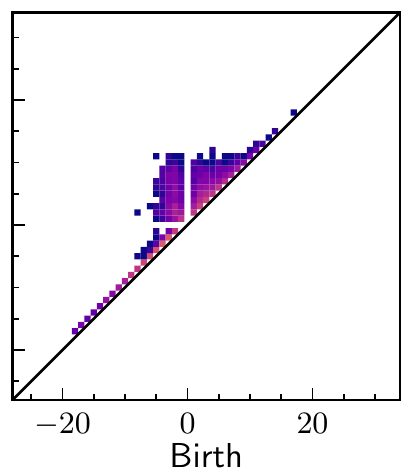}
        \caption{Nickel phase $\mathrm{PD}_1$}
    \end{subfigure}
    \begin{subfigure}{\widthThreeCol\textwidth}
        \centering
        \includegraphics{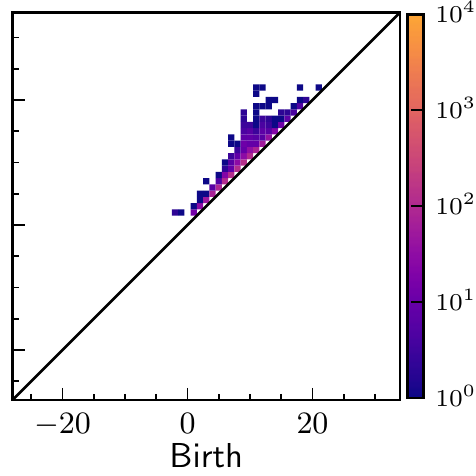}
        \caption{Nickel phase $\mathrm{PD}_2$}
    \end{subfigure}

    \begin{subfigure}{\widthThreeCol\textwidth}
        \centering
        \includegraphics{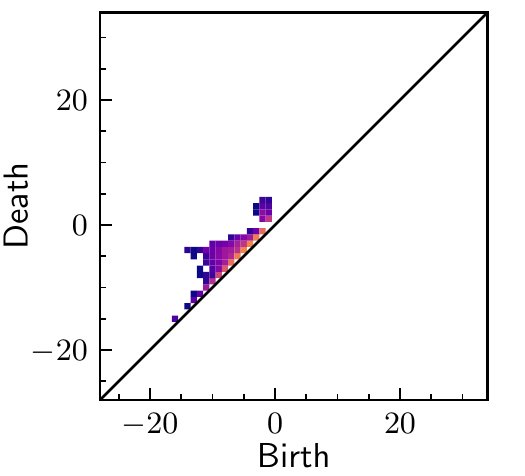}
        \caption{YSZ phase $\mathrm{PD}_0$}
    \end{subfigure}
    \begin{subfigure}{\widthThreeCol\textwidth}
        \centering
        \includegraphics{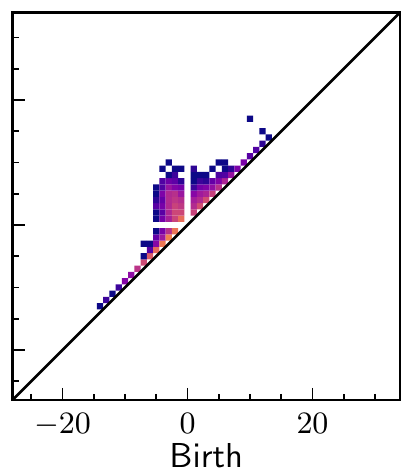}
        \caption{YSZ phase $\mathrm{PD}_1$}
    \end{subfigure}
    \begin{subfigure}{\widthThreeCol\textwidth}
        \centering
        \includegraphics{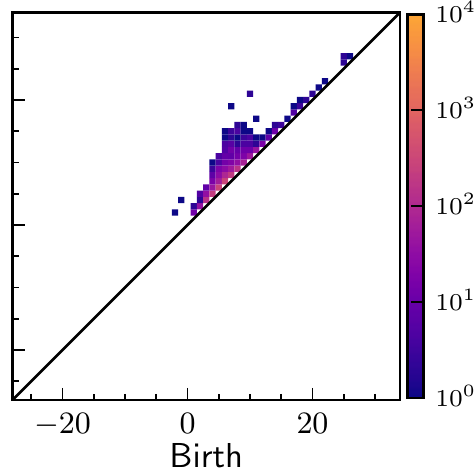}
        \caption{YSZ phase $\mathrm{PD}_2$}
    \end{subfigure}

    \begin{subfigure}{\widthThreeCol\textwidth}
        \centering
        \includegraphics{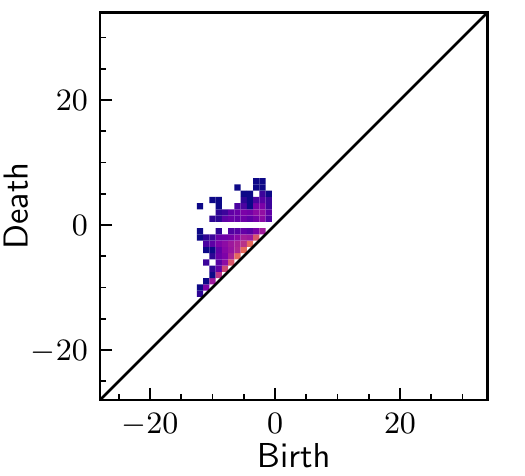}
        \caption{Pore phase $\mathrm{PD}_0$}
    \end{subfigure}
    \begin{subfigure}{\widthThreeCol\textwidth}
        \centering
        \includegraphics{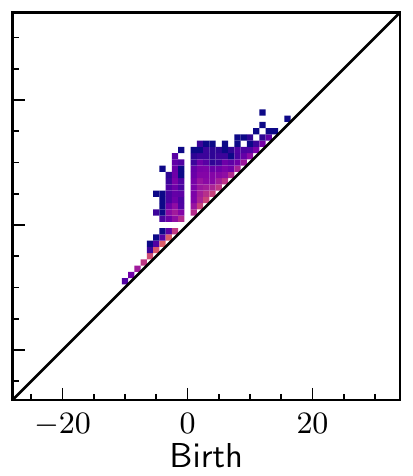}
        \caption{Pore phase $\mathrm{PD}_1$}
    \end{subfigure}
    \begin{subfigure}{\widthThreeCol\textwidth}
        \centering
        \includegraphics{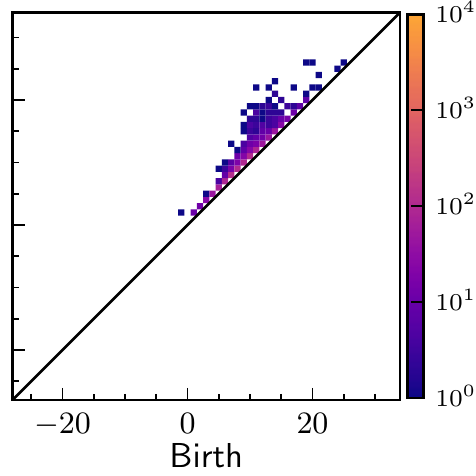}
        \caption{Pore phase $\mathrm{PD}_2$}
    \end{subfigure}

\caption{Persistence diagrams of the three phases of the reference cell}
\label{fig:cell_ref}
\end{figure}

Fig. \ref{fig:cell_ref} shows persistence diagrams $0^{\mathrm{th}}$, $1^{\mathrm{st}}$ and $2^{\mathrm{nd}}$ order of all three phases of the reference cell anode. In the $\mathrm{PD}_0$, a significant difference is observed in the nature of the distribution of birth-death points between the material (nickel and YSZ) and pores. The connection of components is especially valuable from the perspective of confronting the results with data obtained on the basis of the quantitative approach. Continuity of the phase is one of the most important parameters in this method, and the $\mathrm{PD}_0$ representation allows for visual estimation whether the cluster is fully percolated - in this case, the birth-death points on the diagram will have both negative coordinates. For an exemplary brand new cell, it can be concluded that for both nickel and YSZ, the greater number of topological properties, including those most persistent, are in a continuous cluster, while in the case of pores significant amount of points, also far from diagonal, have a positive vertical coordinate. In most cases, $\mathrm{PD}_1$ are similar to each other in terms of the data distribution, although minor differences allow to distinguish the representations of the same phase. Despite minor visual changes, $\mathrm{PD}_1$ are highly hoped for in future research using ANN - the detailed description of the topological features of the channels of individual phases is associated with possibility of improving the mass and energy transport processes taking place in the SOFC anode microstructure. As can be seen in Fig. \ref{fig:cell_ref}, $\mathrm{PD}_2$ for  particular phases is characterized by greater differentiation than cycle counting representations, although some similarities are visible i.e. "stretching" of points of small persistence along the diagonal. $\mathrm{PD}_2$ perfectly represents the scale of material degradation, which will be reflected in the improvement of voids persistence.
\begin{figure}[!t]
\centering
    \begin{subfigure}{0.285\textwidth}
        \centering
        \includegraphics{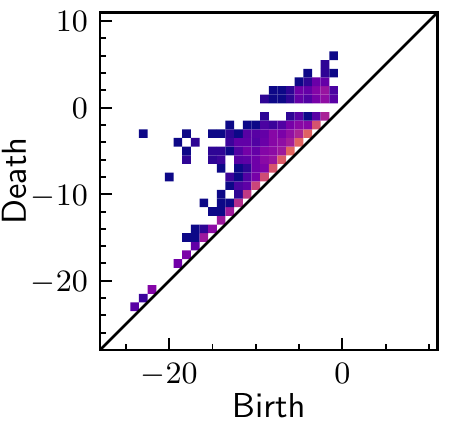}
        \caption{Reference sample}
    \end{subfigure}
    \begin{subfigure}{0.23\textwidth}
        \centering
        \includegraphics{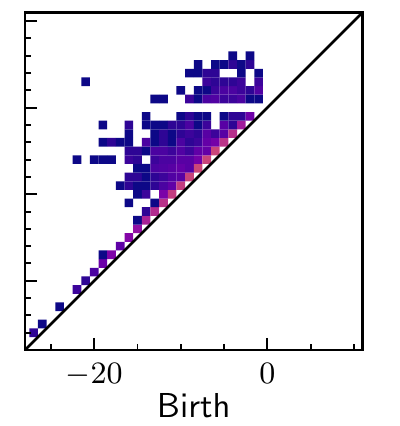}
        \caption{Downstream sample}
    \end{subfigure}
    \begin{subfigure}{0.23\textwidth}
        \centering
        \includegraphics{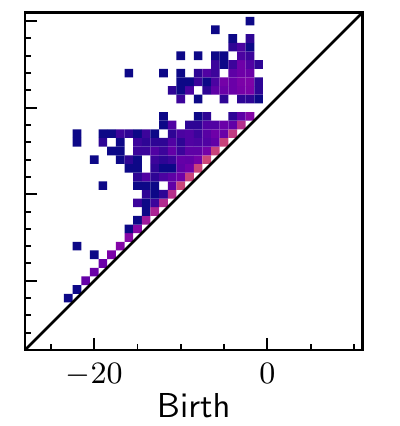}
        \caption{Center sample}
    \end{subfigure}
    \begin{subfigure}{0.23\textwidth}
        \centering
        \includegraphics{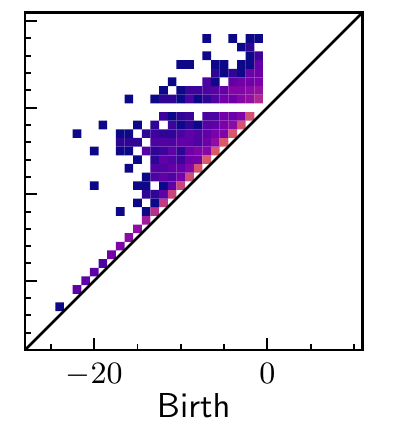}
        \caption{Upstream sample}
    \end{subfigure}
\caption{$\mathrm{PD}_0$ persistence diagrams of the cell 5 for Nickel phase}
\label{cell5_0_Ni}
\end{figure}

To test differences between new and after long-term operation, we have chosen cell no. 5. The selection of the cell is dictated by the possibility of comparing observations based on TDA with data obtained from the quantitative approach methods presented in our previous works \cite{brus2015degradation,brus2015degradationtpb}. $0^{\mathrm{th}}$ order diagrams for Nickel phase are juxtaposed in Fig. \ref{cell5_0_Ni}. In this case, a greater number of features being above 0 in the $Y$ axis are observed in comparison to the representation of the reference sample. When analyzing the microstructure in relation to the sampling site, there is a noticeable trend in the decreasing number of birth-death points with positive vertical coordinate along the fuel flow direction, with an improvement in persistence of the features for diagrams of each sample. 

\begin{figure}[!t]
\centering
\begin{subfigure}{0.285\textwidth}
    \centering
        \includegraphics{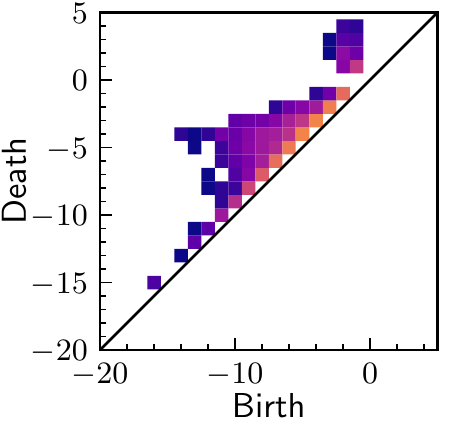}
        \caption{Reference sample}
    \end{subfigure}
    \begin{subfigure}{0.23\textwidth}
        \centering
        \includegraphics{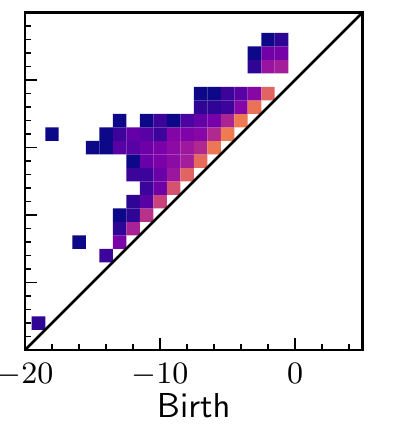}
        \caption{Downstream sample}
    \end{subfigure}
    \begin{subfigure}{0.23\textwidth}
        \centering
        \includegraphics{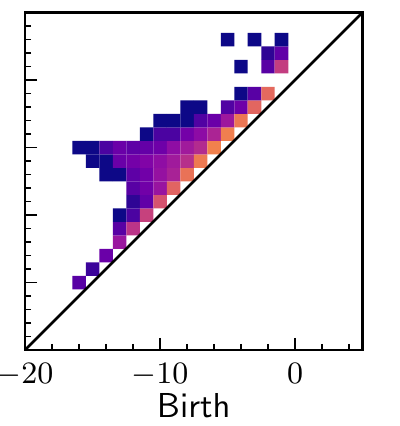}
        \caption{Center sample}
    \end{subfigure}
    \begin{subfigure}{0.23\textwidth}
        \centering
        \includegraphics{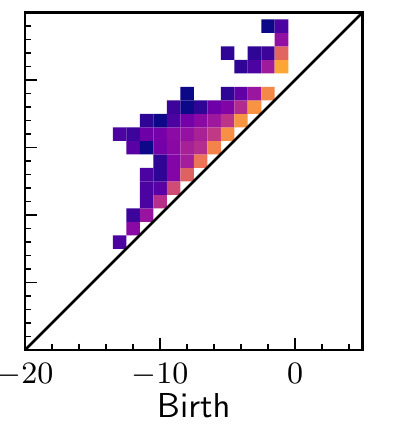}
        \caption{Upstream sample}
    \end{subfigure}
\caption{$\mathrm{PD}_0$ persistence diagrams of the cell 5 for YSZ phase}
\label{cell5_0_YSZ}
\end{figure}

$\mathrm{PD}_0$ diagrams for YSZ are presented in Fig. \ref{cell5_0_YSZ}. The representation of the connected components of YSZ phase of the cell subjected to the degradation process is similar to the corresponding diagram of the reference sample - relatively small differences between individual PDs are observed, but they do not have the specific tendencies. 

\begin{figure}[!t]
\centering
    \begin{subfigure}{0.285\textwidth}
        \centering
        \includegraphics{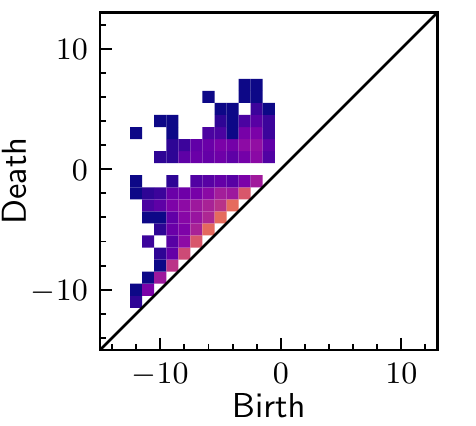}
        \caption{Reference sample}
    \end{subfigure}
    \begin{subfigure}{0.23\textwidth}
        \centering
        \includegraphics{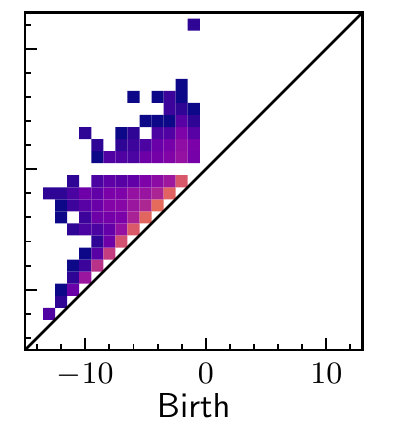}
        \caption{Downstream sample}
    \end{subfigure}
    \begin{subfigure}{0.23\textwidth}
        \centering
        \includegraphics{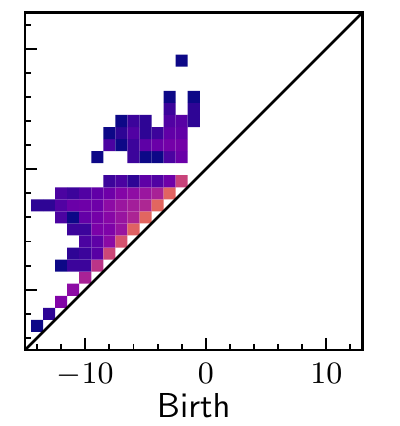}
        \caption{Center sample}
    \end{subfigure}
    \begin{subfigure}{0.23\textwidth}
        \centering
        \includegraphics{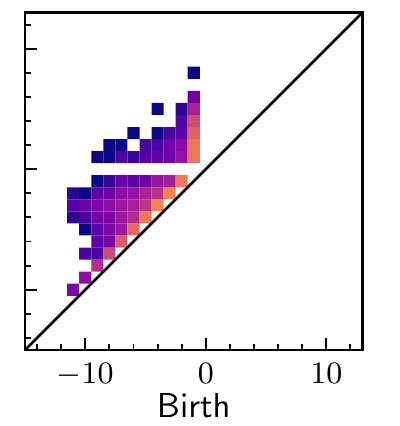}
        \caption{Upstream sample}
    \end{subfigure}
\caption{$\mathrm{PD}_0$ persistence diagrams of the cell 5 for pore phase}
\label{cell5_0_pore}
\end{figure}
The most significant differences in $\mathrm{PD}_0$ of the pore phase, which can be seen on Fig. \ref{cell5_0_pore}, are visible similarly to nickel in the section of the diagram located above zero on the $Y$ axis, with the downstream microstructure resembling reference sample the most. There was no characteristic trend of changes in features in the diagrams for the pore phase depending on the place of sampling. 

It should be emphasized that all $\mathrm{PD}_0$ representations retained the general character of the points distribution typical for the respected phase. Comparing the above observations to the previous analyzes based on the quantitative methods, the tendencies of topological changes for the nickel phase was consistent, while the study of the continuity parameter for samples taken upstream and center sites did not show such significant differences as those presented on the diagram. The share of the YSZ percolated cluster determined on the basis of the quantitative methods was 100\% for both reference sample and all microstructures of the cell at the top of the stack. These data show partial consistence with the observations of the topological properties representations - all samples were similar, while some birth-death points appearance negate the complete phase continuity for each of the microstructures. This difference is caused by the assumption of the continuity of the phase located at the sample boundary. The lowest consistency is found in the data for pores, which in the case of quantitative approach methods indicate an improvement in phase continuity, although indeed the downstream microstructure is the most similar to reference sample. It should be emphasized, however, that the visual assessment of the data contained in $\mathrm{PD}_0$ of the pore phase was inaccurate due to the inconsistency of changes between the individual diagrams.

\begin{figure}[!t]
\centering
     \begin{subfigure}{0.285\textwidth}
        \centering
        \includegraphics{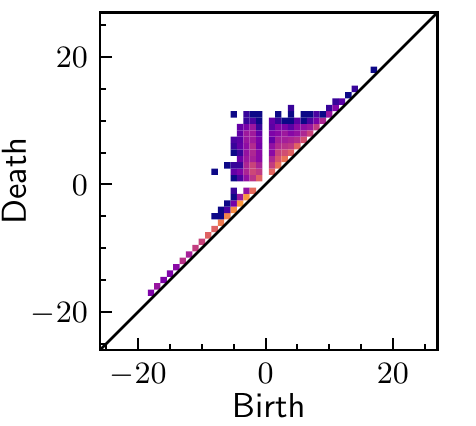}
        \caption{Reference sample}
    \end{subfigure}
    \begin{subfigure}{0.23\textwidth}
        \centering
        \includegraphics{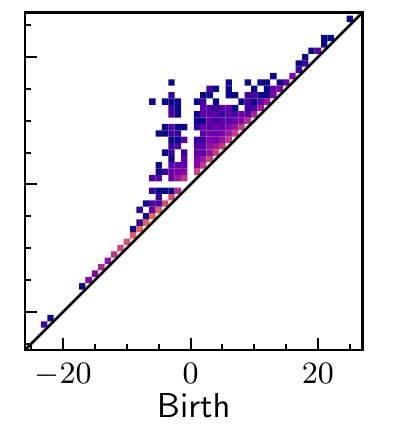}
        \caption{Downstream sample}
    \end{subfigure}
    \begin{subfigure}{0.23\textwidth}
        \centering
        \includegraphics{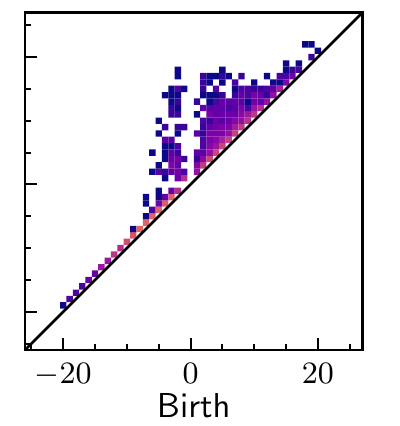}
        \caption{Center sample}
    \end{subfigure}
    \begin{subfigure}{0.23\textwidth}
        \centering
        \includegraphics{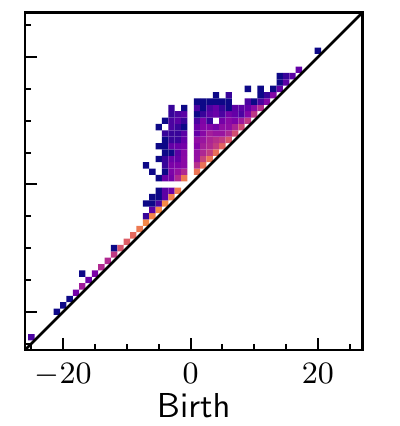}
        \caption{Upstream sample}
    \end{subfigure}
\caption{$\mathrm{PD}_1$ persistence diagrams of the cell 5 for Nickel phase}
\label{cell5_1_Ni}
\end{figure}

PD diagrams of $1^{\mathrm{st}}$ order of nickel phase are presented in Fig. \ref{cell5_1_Ni}. For the diagrams corresponding to the nickel phase, some changes in the trend of the distribution of the birth-death points depending on the place of material collection were observed, while the nature and direction of these changes is difficult to determine. 

\begin{figure}[!t]
\centering
    \begin{subfigure}{0.285\textwidth}
        \centering
        \includegraphics{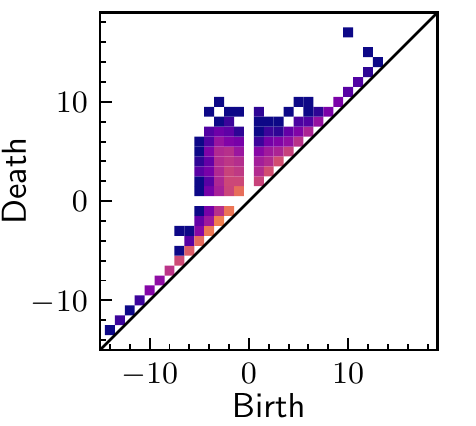}
        \caption{Reference sample}
    \end{subfigure}
    \begin{subfigure}{0.23\textwidth}
        \centering
        \includegraphics{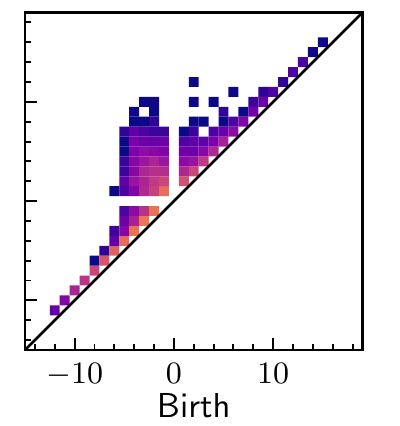}
        \caption{Downstream sample}
    \end{subfigure}
    \begin{subfigure}{0.23\textwidth}
        \centering
        \includegraphics{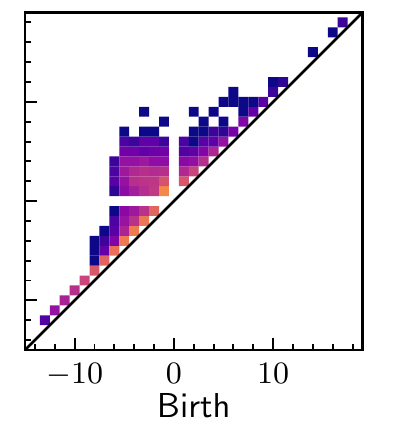}
        \caption{Center sample}
    \end{subfigure}
    \begin{subfigure}{0.23\textwidth}
        \centering
        \includegraphics{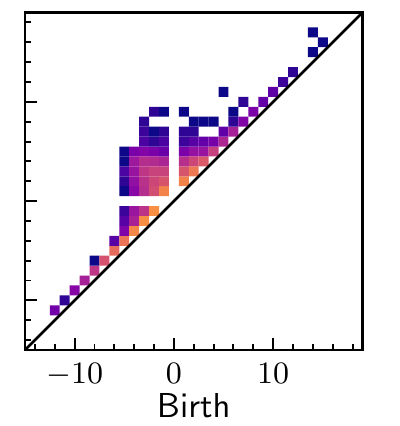}
        \caption{Upstream sample}
    \end{subfigure}
\caption{$\mathrm{PD}_1$ persistence diagrams of the cell 5 for YSZ phase}
\label{cell5_1_YSZ}
\end{figure}

In the case of YSZ, as seen in Fig. \ref{cell5_1_YSZ}, the representations of topological properties are extremely close to each other, and the scale of changes taking place as a result of the degradation process is unnoticeable. 

\begin{figure}[!t]
\centering
    \begin{subfigure}{0.285\textwidth}
        \centering
        \includegraphics{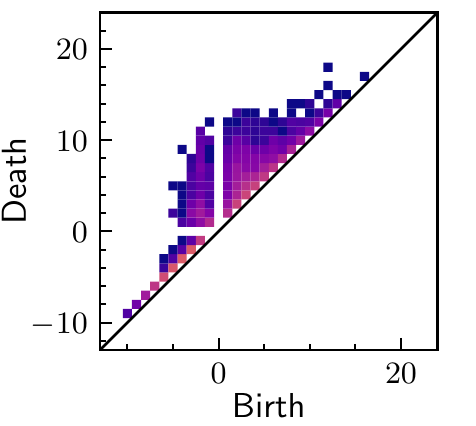}
        \caption{Reference sample}
    \end{subfigure}
    \begin{subfigure}{0.23\textwidth}
        \centering
        \includegraphics{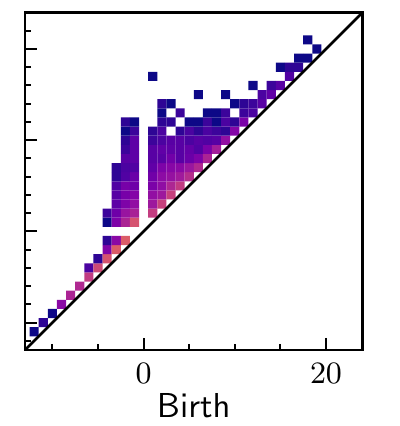}
        \caption{Downstream sample}
    \end{subfigure}
    \begin{subfigure}{0.23\textwidth}
        \centering
        \includegraphics{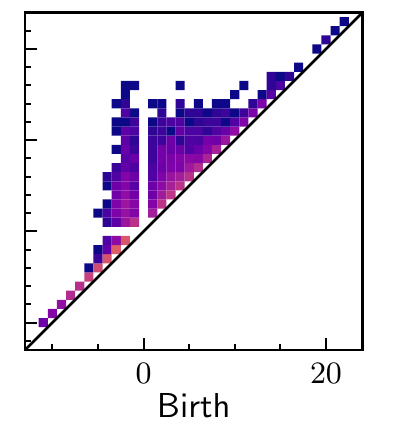}
        \caption{Center sample}
    \end{subfigure}
    \begin{subfigure}{0.23\textwidth}
        \centering
        \includegraphics{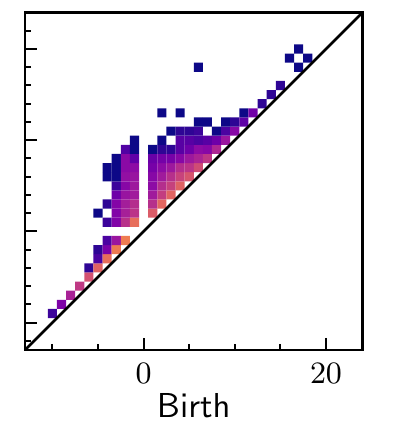}
        \caption{Upstream sample}
    \end{subfigure}
\caption{$\mathrm{PD}_1$ persistence diagrams of the cell 5 for pore phase}
\label{cell5_1_pore}
\end{figure}

Persistent diagrams of 1st order for pore phase are presented in Fig. \ref{cell5_1_pore}. The greatest changes in $\mathrm{PD}_1$ are observed for the pore phase, where in relation to the reference sample there is a greater number of tunnels appearing in further  stages of filtration and this tendency correlates with the direction of fuel flow.

\begin{figure}[!t]
\centering
    \begin{subfigure}{0.285\textwidth}
        \centering
        \includegraphics{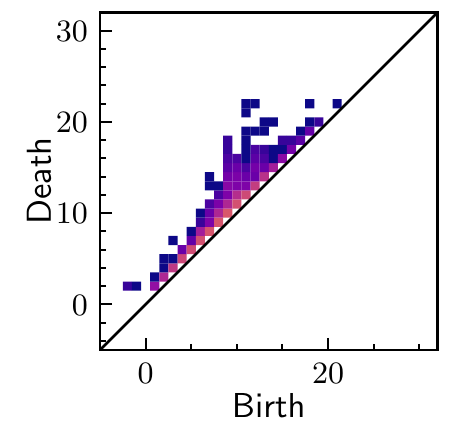}
        \caption{Reference sample}
    \end{subfigure}
    \begin{subfigure}{0.23\textwidth}
        \centering
        \includegraphics{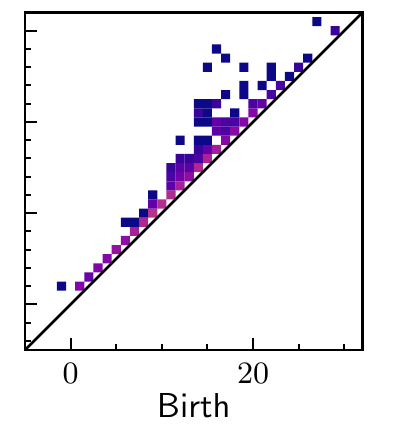}
        \caption{Downstream sample}
    \end{subfigure}
    \begin{subfigure}{0.23\textwidth}
        \centering
        \includegraphics{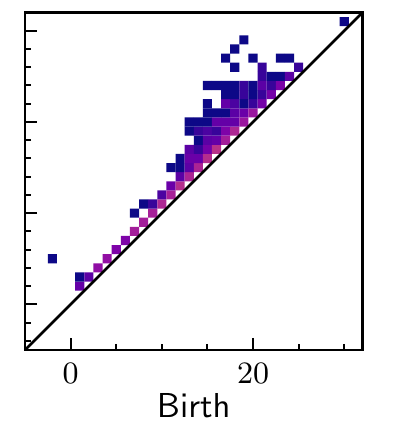}
        \caption{Center sample}
    \end{subfigure}
    \begin{subfigure}{0.23\textwidth}
        \centering
        \includegraphics{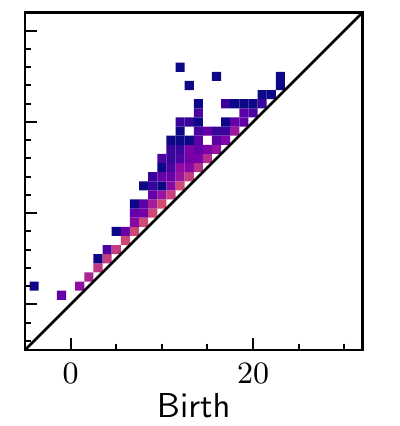}
        \caption{Upstream sample}
    \end{subfigure}
    \caption{$\mathrm{PD}_2$ persistence diagrams of the cell 5 for Nickel phase}
\label{cell5_2_Ni}
\end{figure}

The $2^{\mathrm{nd}}$ order diagrams of Nickel phase are presented on Fig. \ref{cell5_1_Ni}. Topological changes in the microstructure due to material degradation are much more visible in the case of $\mathrm{PD}_2$. The diagrams representing the changes in the nickel phase show a significant increase in the number of birth-death points with a later birth time in comparison to the reference sample and these properties are characterized by relatively high persistence. There was no characteristic tendencies in the number and distance of points from the diagonal depending on the place of sampling.
\begin{figure}[!t]
\centering
    \begin{subfigure}{0.285\textwidth}
        \centering
        \includegraphics{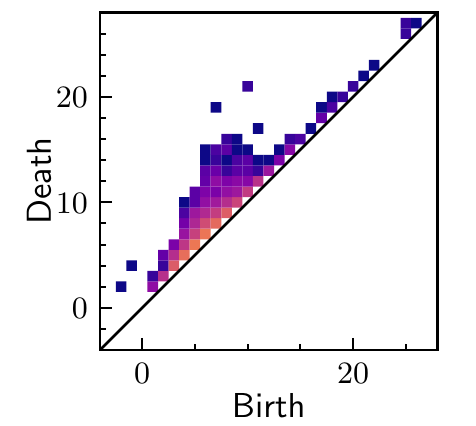}
        \caption{Reference sample}
    \end{subfigure}
    \begin{subfigure}{0.23\textwidth}
        \centering
        \includegraphics{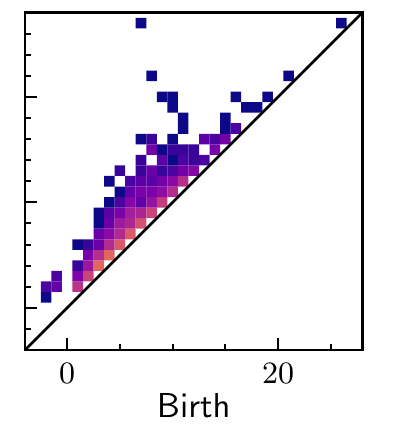}
        \caption{Downstream sample}
    \end{subfigure}
    \begin{subfigure}{0.23\textwidth}
        \centering
        \includegraphics{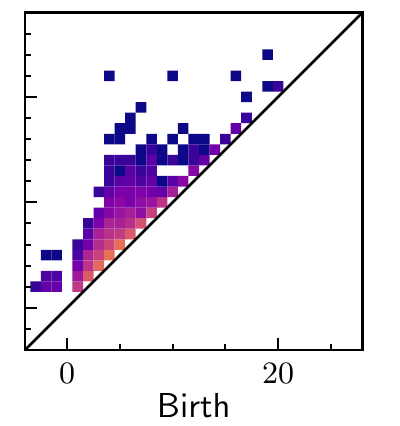}
        \caption{Center sample}
    \end{subfigure}
    \begin{subfigure}{0.23\textwidth}
        \centering
        \includegraphics{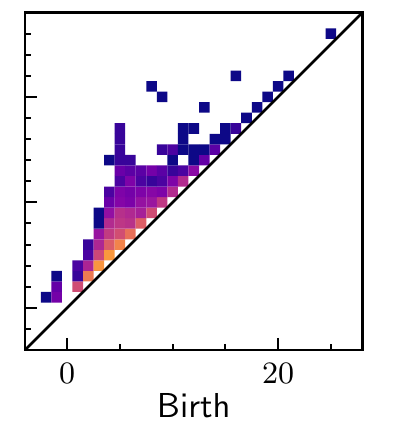}
        \caption{Upstream sample}
    \end{subfigure}
    \caption{$\mathrm{PD}_2$ persistence diagrams of the cell 5 for YSZ phase}
\label{cell5_2_YSZ}
\end{figure}

$\mathrm{PD}_2$ of the YSZ phase is characterized by a much greater number of the presented features than for nickel, as shown in Fig. \ref{cell5_2_YSZ}. Comparing the diagrams of the microstructure subjected to the aging process with the material from the brand new cell, a general increase in the number of birth-death points and appearance of a relatively large group of individual properties with very high persistence, which were born in the final stages of filtration, are observed. 

\begin{figure}[!t]
\centering
     \begin{subfigure}{0.285\textwidth}
        \centering
        \includegraphics{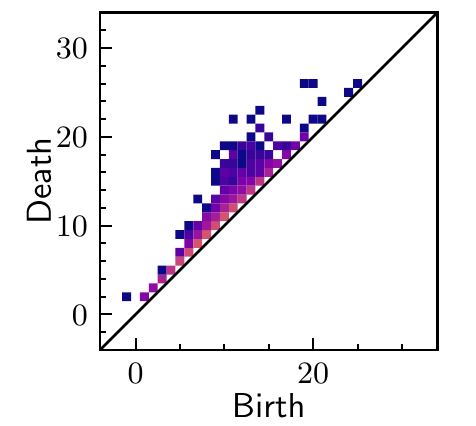}
        \caption{Reference sample}
    \end{subfigure}
    \begin{subfigure}{0.23\textwidth}
        \centering
        \includegraphics{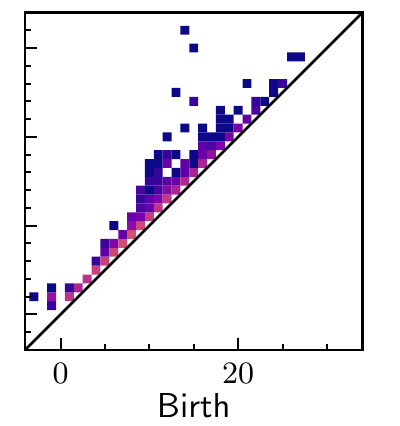}
        \caption{Downstream sample}
    \end{subfigure}
    \begin{subfigure}{0.23\textwidth}
        \centering
        \includegraphics{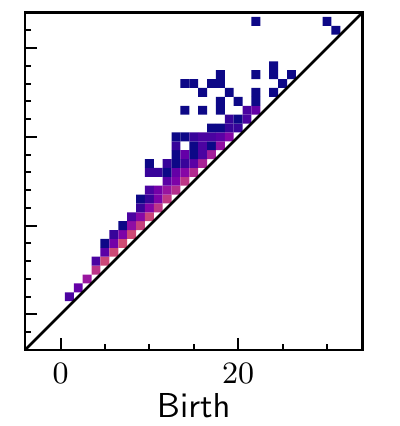}
        \caption{Center sample}
    \end{subfigure}
    \begin{subfigure}{0.23\textwidth}
        \centering
        \includegraphics{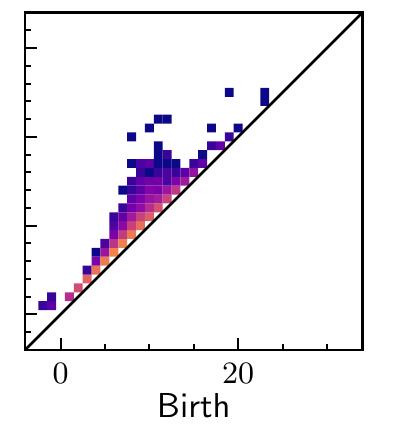}
        \caption{Upstream sample}
    \end{subfigure}
\caption{$\mathrm{PD}_2$ persistence diagrams of the cell 5 for pore phase}
\label{cell5_2_Pore}
\end{figure}
For the $\mathrm{PD}_2$ representation of the pore phase, shown in Fig. \ref{cell5_2_Pore}, a certain trend in the distribution of topological properties can be noticed depending on the place of sampling. Starting from the reference sample and then carrying out the analysis of microstructures in the direction of the the fuel flow, the "flattening" along the diagonal of the main grouping of birth-death points is observed, with a simultaneous increase in single points of high persistence for each subsequent sample. 

\section{Conclusions}
The article used computational topology to analyze microstructural data from solid oxide fuel cells before and after long-term operation. It is the first report that uses persistent homology for microstructure analysis of solid oxide fuel cell electrodes. Summarizing the presented observations, it can be concluded that even a strongly imperfect visual analysis of topological data allows for the initial selection of information about microstructure and distinguishing trends of topological changes taking place in the material. In the case of $\mathrm{PD}_1$ and $\mathrm{PD}_2$ analyzes, there is no benchmark, which for $\mathrm{PD}_0$ were the data obtained using the quantitative methods. However, it should be emphasized that ultimately the topological data analysis methods should cooperate with deep-learning algorithms, which would allow the identification of characteristic changes impossible to observe by the human eye. 

\section*{Acknowledgments}
Research project supported by the program "Excellence Initiative – Research University" for the AGH University of Science and Technology.

\bibliographystyle{unsrtnat}
\bibliography{references} 

\end{document}